\documentclass[a4paper,11pt]{article}
\pdfoutput=1 

\usepackage{jcappub} 

\usepackage[T1]{fontenc} 
\usepackage{ifthen}
\usepackage{epsfig}
\usepackage{booktabs} 
\usepackage{natbib}
\usepackage{bbold}
\usepackage{diagbox}
\usepackage{slashed}

\usepackage{tikz-feynman}
\usetikzlibrary{external}
\immediate\write18{mkdir -p pgf-img}
\usepackage{capt-of}

\newcommand{\erf}{\mathop{\mathrm{erf}}}
\newcommand{\beqra}{\begin{flalign}}
\newcommand{\eeqra}{\end{flalign}}
\newcommand{\beq}{\begin{equation}}
\newcommand{\eeq}{\end{equation}}

\newcommand{\vs}{\vec{s}}
\newcommand{\vS}{\vec{S}}

\newcommand{\vt}{\vec{v}^{\bot}}
\newcommand{\sn}{\vec{S}_{N}^{\,r'r}}
\newcommand{\sx}{\vec{S}_{\chi}^{\,s's}}
\newcommand{\sX}{\vec{S}_{X}^{\,s's}}
\newcommand{\sndag}{\vec{S}_{N}^{\,rr'}}
\newcommand{\sxdag}{\vec{S}_{\chi}^{\,ss'}}
\newcommand{\sXdag}{\vec{S}_{X}^{\,ss'}}
\newcommand{\ubar}{\bar{u}}

\newcommand{\Sy}{\mathcal{S}}

\newcommand{\es}{\vec{e}_s}
\newcommand{\esp}{\vec{e}\,'_{s'}}
\newcommand{\lvector}[2]{\begin{pmatrix}
#1\\
#2
\end{pmatrix}}

\newcommand{\niM}{\mathcal{M}}

\usepackage{ccicons} 
\usepackage{listings}
\usepackage{color}   
\usepackage{braket}  
\usepackage{hyperref}
\usepackage{dsfont}  
\usepackage{leftidx} 
\RequirePackage{mathrsfs}
\RequirePackage{dsfont}
\usepackage{bm}
\usepackage{placeins}
\usepackage{float}
\allowdisplaybreaks

\title{\boldmath Direct detection of fermionic and vector dark matter with polarised targets}
\author[a]{Riccardo Catena}
\author[a]{K\aa re Fridell}
\author[a,b,c]{and Vanessa Zema}
\affiliation[a]{Chalmers University of Technology, Department of Physics, SE-412 96 G\"oteborg, Sweden}
\affiliation[b]{GSSI-Gran Sasso Science Institute, 67100, L'Aquila, Italy}
\affiliation[c]{INFN, Laboratori Nazionali del Gran Sasso, I-67010 Assergi, Italy}
\emailAdd{catena@chalmers.se}
\emailAdd{karef@chalmers.se}
\emailAdd{vanessa.zema@chalmers.se}

\abstract{We study the scattering of Milky Way dark matter (DM) particles by spin-polarised target nuclei within a set of simplified models for fermionic and vector DM where DM interacts with spin 1/2 point-like nuclei through the exchange of a vector or pseudo-vector mediator particle.~This study is motivated by the possibility of using polarised targets to gain novel insights into the nature of DM.~For fermionic DM, we provide an explicit expression for the polarised DM-nucleus scattering cross section refining previous results found in the literature.~For vector DM, we calculate the polarised cross section for DM-nucleus scattering for the first time.~We find that polarised targets can in principle be used to discriminate fermionic from vector DM.}

\begin{document}
\maketitle
\flushbottom

\section{Introduction}
\label{sec:introduction}
There is strong evidence for the presence of invisible mass, or dark matter (DM), in our Universe~\cite{Bertone:2004pz}.~While the evidence for DM is strong, it is only indirect and based upon gravitational effects on, e.g., stars, galaxies, galaxy clusters, the cosmic microwave background radiation, and the Universe at large~\cite{Jungman:1995df}.~For this reason, the nature of DM remains a mystery.~In astroparticle physics it is standard to assume that DM is made of hypothetical particles, whose present cosmological density is set in the early Universe by chemical decoupling from the thermal bath~\cite{Bertone:2016nfn}.~WIMPs (for Weakly Interacting Massive Particles) represent the most extensively studied realisation of this hypothesis~\cite{Arcadi:2017kky}.~So far, WIMPs escaped detection, but complementary strategies have been proposed in the past decades to detect WIMPs in the laboratory, in space, or at particle accelerators~\cite{Roszkowski:2017nbc}.~The so-called direct detection technique~\cite{Drukier:1983gj,Goodman:1984dc} will be crucial in testing the WIMP hypothesis in the coming years~\cite{Baudis:2012ig,Catena:2014epa,Catena:2014hla,Catena:2015vpa}.~It primarily searches for non-relativistic DM-nucleus scattering events in low-background experiments located deep underground~\cite{Lewin:1995rx}.

Assessing the performance of a direct detection experiment, key parameters are the exposure, the energy threshold, and the sensitivity to the direction of nuclear recoils~\cite{Undagoitia:2015gya}.~While significant progress has been made along the three directions in the past decade, only recently it has been realised that the use of polarised targets would provide an additional handle on the nature of DM that direct detection experiments can exploit~\cite{Chiang:2012ze}.~For example, it has been found that a polarised $^{3}$He detector would in principle be able to discriminate DM signal events from solar neutrino-induced nuclear recoils with an efficiency of 98\%, when the orientation of the polarisation axis is antiparallel to the direction of the Sun~\cite{Franarin:2016ppr}.~$^{3}$He is an interesting target also for other reasons, e.g.:~it has no intrinsic X-ray emission; it has a low natural radioactive background; and rejection of neutron-induced background events can easily be achieved through the process n+$^3$He$\rightarrow$p+3H+764 keV~\cite{Amarian:2002ar,Amarian:2003jy}.~Methods to produce large samples of spin-polarised noble gasses, including helium, argon and xenon, are reviewed in~\cite{Bouchiat:1960dsd,Walker:1997zzc}.

The rate of DM scattering by polarised nuclei has for the first time been computed in~\cite{Chiang:2012ze}, focusing on fermionic DM scattering off a point-like spin 1/2 nucleus, and on interactions mediated by a vector or pseudo-vector mediator.~A related study is the one presented in \cite{Franarin:2016ppr}, which focuses on polarised neutrino-nucleus scattering events as an experimental background to direct DM searches.~In this article, we refine the results presented in~\cite{Chiang:2012ze} for fermionic DM (by correcting two sign errors found in the scattering cross section in Eq.~(2) of~\cite{Chiang:2012ze}), and extend them to the case of vector DM.~In the case of scalar DM-nucleus scattering~\cite{Brod:2017bsw,Bishara:2016hek}, the leading nuclear polarisation dependent terms arise at third order in the momentum transfer and are therefore expected to be generically smaller than the ones arising from models for fermionic and vector DM~\cite{Chiang:2012ze}.~Our study is motivated by the possibility of using polarised targets to gain novel insights into the nature of DM which would otherwise be difficult to extract from the result of operating direct detection experiments.~For example, we find that a signal at polarised direct detection experiments could in principle contain information on the DM particle spin.

The article is organised as follows.~In Sec.~\ref{sec:theory} we review the theory of polarised DM-nucleus scattering, focusing on nuclear recoil energy spectra and angular distributions which can be measured at direct detection experiments.~Emphasis will be placed on the relation between these quantities and the squared modulus of the amplitude for polarised DM-nucleus scattering.~Sec.~\ref{sec:amplitudes} provides explicit expressions for the latter, focusing on a general set of simplified models for fermionic and vector DM~\cite{Catena:2017xqq,Baum:2017kfa,Catena:2017wzu,Dent:2015zpa}, assuming a point-like spin 1/2 nucleus as a target.~For all DM models considered here, nuclear recoil energy spectra and angular distributions are numerically evaluated in Sec.~\ref{sec:results}, where we also comment on the dependence of our results on the DM spin.~We conclude in Sec.~\ref{sec:conclusion}, collect useful Feynman rules for vector DM in Appendix~\ref{sec:rules}, and present an independent derivation of our results for fermionic DM in Appendix~\ref{sec:vector_v2}.

\section{Polarised dark matter-nucleus scattering}
\label{sec:theory}
\begin{figure}[t]
\begin{center}
\begin{minipage}[t]{0.49\linewidth}
\centering
\includegraphics[width=0.8\textwidth]{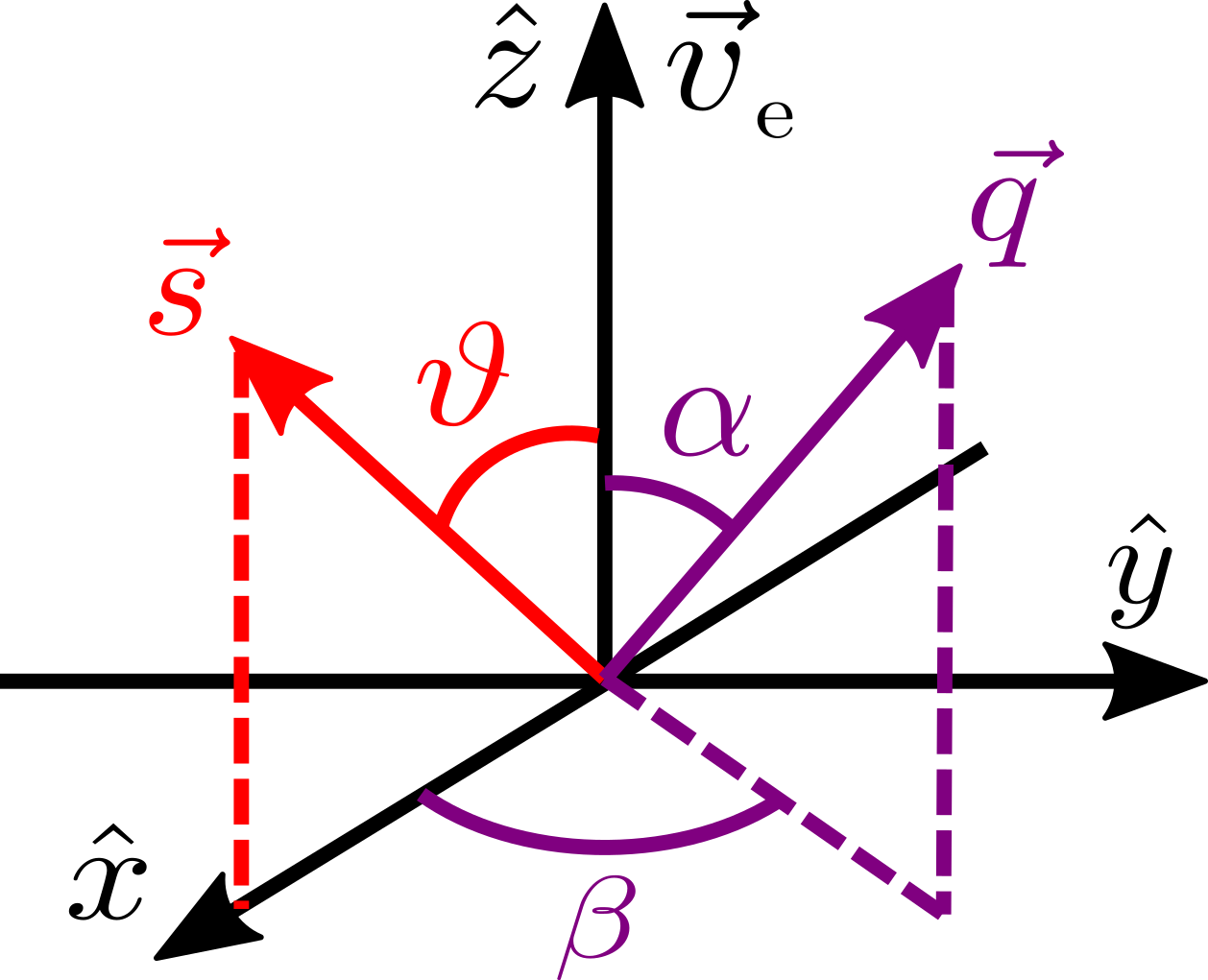}
\end{minipage}
\begin{minipage}[t]{0.49\linewidth}
\centering
\includegraphics[width=0.8\textwidth]{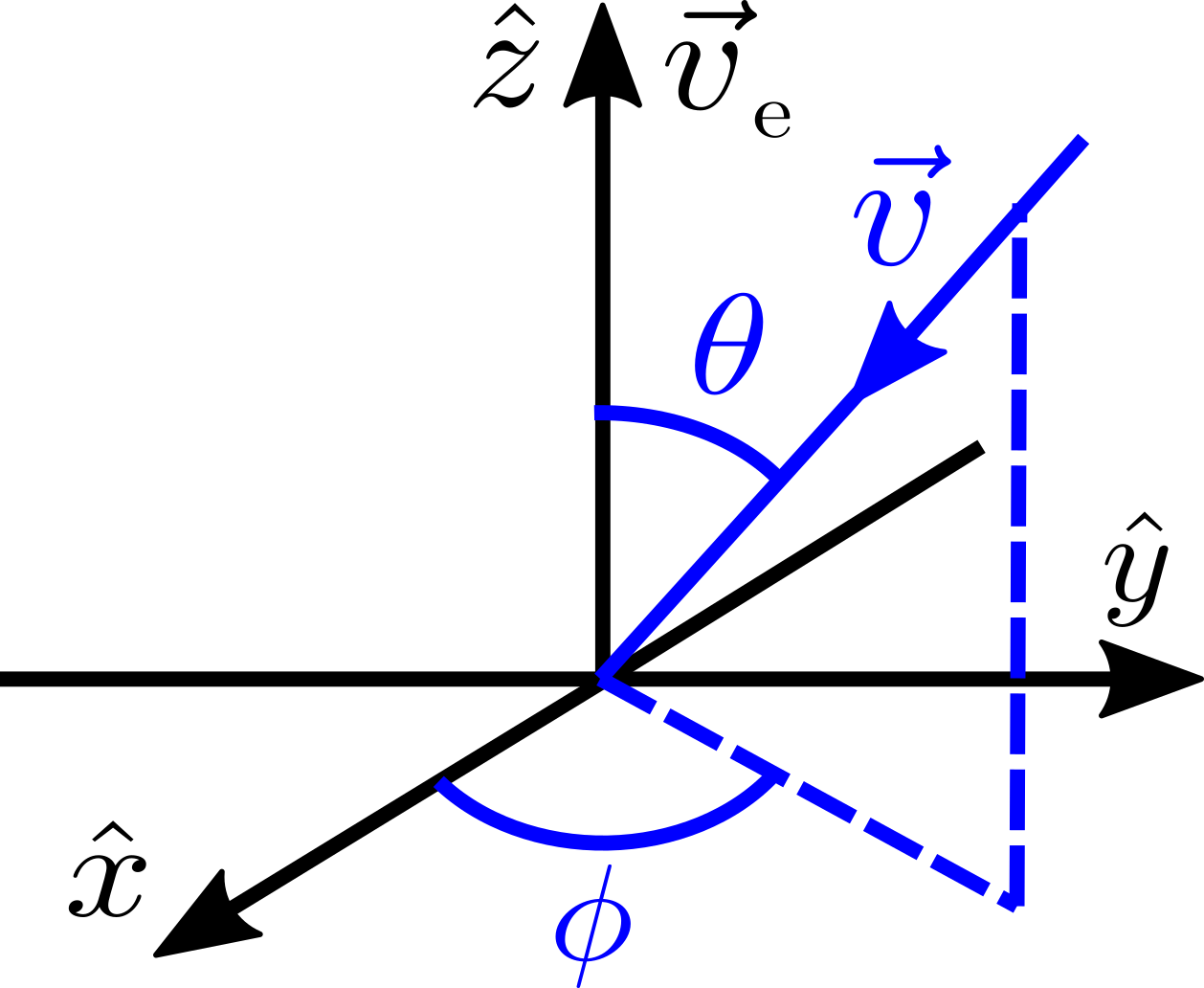}
\end{minipage}
\end{center}
\caption{An illustration of the coordinate system that is used to evaluate the triple differential rate of DM nucleus scattering events, Eq.~(\ref{eq:triple_diff_event_rate_spherical3}).~The red arrow identifies the direction of the polarisation vector of the target nuclei, which without loss of generality is taken to be in $xz$-plane.~The purple arrow corresponds to the direction of nuclear recoil, and the blue arrow to the direction of the incoming DM particle.~By construction, the velocity of the Earth in the galactic rest frame is taken to be in the $z$-direction.}  
\label{fig:frame}
\end{figure}

In this section we review the theory of polarised DM-nucleus scattering.~We are interested in computing nuclear recoil energy spectra and angular distributions resulting from the scattering of DM particles by a sample of spin-polarised nuclei, and which can in principle be measured at direct detection experiments.~To this end, let us introduce the polarisation vector of the target nuclei, $\vs = 2\vS_N$, where $\vS_N=\vec{\sigma}_N/2$ is the nuclear spin operator, and $\vec{\sigma}_N$ is a tridimensional vector whose components are Pauli matrices acting on the nucleus spin space.~If DM couples to nuclei via parity violating interactions, the cross section for polarised DM-nucleus scattering can explicitly depend on $\vs$~\cite{Chiang:2012ze}.~In this case, the vector $\vs$ breaks the azimuthal symmetry of DM-nucleus scattering, and it is convenient to construct physical observables in terms of the triple differential rate of DM-nucleus scattering events per unit detector mass,
\begin{equation}
\label{eq:triple_diff_event_rate}
\frac{{\rm d}R}{{\rm d}E_{\text{R}}{\rm d}\Omega} = \frac{\rho_\chi}{m_\chi m_N}\int {\rm d}^3v\, vf(\vec{v}) \frac{{\rm d}\sigma}{{\rm d}E_{\text{R}}{\rm d}\Omega}\,,
\end{equation}
where $\rho_\chi\simeq~0.4$~GeV~cm$^{-3}$~\cite{Sivertsson:2017rkp,Bozorgnia:2013pua,Catena:2011kv,Catena:2009mf}, $m_\chi$ is the DM particle mass, $f(\vec{v})$ and $\vec{v}$ the DM velocity distribution and incoming velocity in the detector rest frame, respectively, and ${\rm d}\sigma/{\rm d}E_{\text{R}}/{\rm d}\Omega$ the triple differential scattering cross section for polarised DM-nucleus scattering, which can be expressed as follows~\cite{Gondolo:2002np}
\begin{equation}
\label{eq:diff_cross_section2}
\frac{{\rm d}\sigma}{{\rm d}E_R{\rm d}\Omega} = \frac{v}{2\pi}\frac{{\rm d}\sigma}{{\rm d}E_R}\delta\left(\vec{v}\cdot\hat{q} - \frac{q}{2\mu}\right).
\end{equation}
Here, $q$ is the momentum transferred along the nuclear recoil direction $\hat{q}$, ${\rm d}\Omega$ is an infinitesimal solid angle around $\hat{q}$, and the differential cross section for polarised DM-nucleus scattering is given by~\cite{DelNobile:2013sia,Catena:2014uqa,Catena:2015uua}
\begin{equation}
\label{eq:diff_cross_section1}
\frac{{\rm d}\sigma}{{\rm d}E_R} = \frac{1}{32\pi}\frac{1}{m_\chi^2 m_N}\frac{1}{v^2}|\overline{\niM}|^2,
\end{equation}
where $E_R = q^2/(2m_N)$ is the nuclear recoil energy, $m_N$ the target nucleus mass, and $|\overline{\niM}|^2$ the squared modulus of the amplitude for DM-nucleus scattering, summed over the final DM and nucleus spin states, and averaged over the initial DM spin configurations.~Here, the initial nucleus polarisation state is assumed to be known (and the same for all nuclei in the detector).~For the velocity distribution in Eq.~(\ref{eq:triple_diff_event_rate}), we assume a Maxwell-Boltzmann distribution truncated at the galactic escape velocity $v_{\text{esc}}=544$~km~s$^{-1}$, and with most probable speed given by the circular speed of the local standard of rest, $v_0=220$~km~s$^{-1}$.~More specifically, 
\begin{align}
\label{eq:velodis}
f(\vec{v}) &= \frac{1}{N}e^{-(\vec{v} + \vec{v}_{\text{e}})^2/v_0^2}\,; \qquad |\vec{v} + \vec{v}_{\text{e}}|\le v_{\rm esc}  \,,
\end{align}
where
\begin{equation}
N = \pi v_0^2\left[\sqrt{\pi}v_0\erf\left(\frac{v_{\text{esc}}}{v_0}\right) - 2v_{\text{esc}}e^{-v_{\text{esc}}^2/v_0^2}\right]\,,
\end{equation}
and $\vec{v}_{\text{e}}$ is the Earth velocity in the galactic rest frame.~An explicit expression for $\vec{v}_{\text{e}}$ as a function of time is given in~\cite{Lee:2013xxa}.~For simplicity, here we assume for $\vec{v}_{\text{e}}$ the constant magnitude 232~km~s$^{-1}$~\cite{Freese:2012xd}.~Taking into account the finite value of the escape velocity $v_{\text{esc}}$, the tridimensional velocity integral in Eq.~(\ref{eq:triple_diff_event_rate}) must be divided into two parts, as explicitly shown below
\begin{align}
\int {\rm d}^3v &= \int_{v_{\text{min}}}^{v_\text{esc}-v_{\text{e}}} {\rm d}v\,v^2 \int_{-1}^{+1} {\rm d}\hspace{-0.04 cm}\cos\hspace{-0.04 cm}\theta \int_0^{2\pi} {\rm d} \phi \nonumber\\ &+ \int_{v_\text{esc}-v_{\text{e}}}^{v_\text{esc}+v_{\text{e}}} {\rm d}v\,v^2 \int_{-1}^{\frac{v_\text{esc}^2-v^2-v^2_{\text{e}}}{2v v_{\text{e}}}} {\rm d}\hspace{-0.04 cm}\cos\hspace{-0.04 cm}\theta \int_0^{2\pi} {\rm d} \phi \,,
\end{align}
where $v_{\rm min}=\sqrt{m_N E_{R}/(2 \mu^2)}$ is the minimum velocity a DM must have in the detector rest frame to deposit a nuclear recoil energy $E_R$ in the scattering, and $\mu$ is the DM-nucleus reduced mass.~Furthermore, the integral over ${\rm d} v$ in Eq.~(\ref{eq:triple_diff_event_rate}) can be performed analytically by using the identity  
\begin{equation}
\delta\left(\vec{v}\cdot\hat{q} - \frac{q}{2\mu}\right) = \frac{\delta(v-\bar{v})}{|\hat{v}\cdot\hat{q}|} \,,
\label{eq:deltas}
\end{equation}
where $\bar{v}=q/[2\mu (\hat{v}\cdot\hat{q})]$, and $\hat{v}$ is a unit vector in the direction of the incoming DM velocity, $\vec{v}=v\hat{v}$.~Using Eq.~(\ref{eq:deltas}), we find the following expression fo the triple differential rate defined in Eq.~(\ref{eq:triple_diff_event_rate})
\begin{equation}
\label{eq:triple_diff_event_rate_spherical3}
\frac{{\rm d}R}{{\rm d}E_{\text{R}}{\rm d}\Omega} = \frac{\rho_\chi}{64\pi^2Nm_\chi^3m_N^2}\sum_{\ell=1}^2\int_{-1}^{+1}{\rm d}\hspace{-0.04 cm}\cos\hspace{-0.04 cm}\theta\int_0^{2\pi}{\rm d}\phi \frac{\bar{v}^2}{|\hat{v}\cdot\hat{q}|}
e^{-(\bar{v}^2 + v_{\text{e}}^2+2 \bar{v} v_e \cos\hspace{-0.04 cm}\theta )/v_0^2}
\,|\overline{\niM}|^2 \, \Theta_\ell\,,
\end{equation}
where $\Theta_1=\Theta(\bar{v}-v_{\rm min}) \Theta((v_{\rm esc}-v_{\rm e})-\bar{v})$ and $\Theta_2=\Theta((v_\text{esc}^2-\bar{v}^2-v^2_{\text{e}})/(2\bar{v} v_{\text{e}})-\cos\theta)\Theta(\bar{v}-(v_{\rm esc}-v_{\rm e}))\Theta((v_{\rm esc}+v_{\rm e})-\bar{v})$.~In order to evaluate Eq.~(\ref{eq:triple_diff_event_rate_spherical3}), we choose a reference frame with $z$-axis along the direction of $\vec{v}_{\rm e}$ (see Fig.~\ref{fig:frame}).~In this frame, the vectors $\hat{q}$, $\hat{v}$, $\hat{s}=\vec{s}/|\vec{s}\,|$ and $\hat{v}'=\vec{v}\,'/|\vec{v}\,'|$, where $\vec{v}\,'$ is the DM particle velocity after scattering, can be written as follows
\begin{align}
\hat{q} &= (\sin\alpha\cos\beta, \sin\alpha\sin\beta, \cos\alpha)\,,\nonumber\\
\hat{v} &= (\sin\theta\cos\phi, \sin\theta\sin\phi, \cos\theta)\,,\nonumber\\
\hat{s} &= (\sin\vartheta, 0, \cos\vartheta)\,,\nonumber\\
\hat{v}' &= \frac{v}{v'}\hat{v} + \frac{q}{m_\chi v'}\hat{q}\,.
\end{align}
and the infinitesimal solid angle ${\rm d}\Omega$ can consistently be expressed in terms of the angles $\alpha$ and $\beta$, i.e.~${\rm d}\Omega={\rm d}\hspace{-0.04 cm}\cos\hspace{-0.04 cm}\alpha\,{\rm d}\beta$.

\section{Scattering amplitudes}
\label{sec:amplitudes}
In this section we calculate the squared modulus of the amplitude for polarised DM-nucleus scattering, $|\overline{\niM}|^2$, in simplified models for fermionic and vector DM where the DM-nucleus interaction is mediated by the exchange of a heavy spin 1 particle.~Here, a bar denotes average over initial DM spin states, and sum over final DM and nucleus spin configurations.~Below, incoming and outgoing four-momenta for the DM particle (target nucleus) are denoted by $p$ and $p'$ ($k$ and $k'$), respectively.~Furthermore, we define as $\vec{q}=\vec{k}-\vec{k}\,'=\vec{p}\,'-\vec{p}$ the momentum transferred from the nucleus to the DM particle~\cite{Fitzpatrick:2012ix,Anand:2013yka}.~We also introduce the transverse relative velocity vector,
\begin{align}
\label{eq:vt}
2\vt = \vec{v} + \vec{v}\,' + \frac{m_\chi}{m_N}(\vec{v}\,' - \vec{v})\,.
\end{align}
Finally, we assume nuclei to be point-like, with spin 1/2, and in a given initial spin state.~The calculation of $|\overline{\niM}|^2$ will allow us to numerically evaluate the cross section in Eq.~(\ref{eq:diff_cross_section1}) and the rate in Eq.~(\ref{eq:triple_diff_event_rate_spherical3}).

\subsection{Fermionic dark matter}
In the case of fermionic DM, we calculate $|\overline{\niM}|^2$ assuming the following Lagrangian for DM-nucleus interactions
\begin{align}
\label{eq:spin12_lag_full}
\mathscr{L} & = i \bar{\chi}\slashed D\chi - m_{\chi}\bar{\chi}\chi - \frac{1}{4}\mathcal{G}_{\mu\nu}\mathcal{G}^{\mu\nu} + \frac{1}{2}m_G^2G_\mu G^\mu \nonumber\\ 
&+ i\bar{N}\slashed DN - m_{N}\bar{N}N \nonumber\\
&- \lambda_3\bar{\chi}\gamma^\mu\chi G_\mu - \lambda_4\bar{\chi}\gamma^\mu\gamma_5\chi G_\mu \nonumber\\
&- h_3\bar{N}\gamma^\mu NG_\mu - h_4\bar{N}\gamma^\mu\gamma_5 NG_\mu  \,,
\end{align}
where $\chi$ and $N$ are the DM particle and target nucleus spinor fields, respectively, $G_\mu$ is a vector field describing the particle that mediates the interactions between DM and nuclei, $\mathcal{G}_{\mu\nu}$ is the associated field strength tensor, $\gamma^\mu$, $\mu=0,1,2,3$, and $\gamma_5$ are the five gamma matrices, and $\lambda_3$, $\lambda_4$, $h_3$ and $h_4$ are real coupling constants.~In Eq.~(\ref{eq:spin12_lag_full}), $\slashed D$ represents a gauge covariant derivative, and a bar corresponds to hermitian conjugation times $\gamma^0$.~Finally, nucleus mass, DM particle mass, and mediator mass are denoted by $m_N$, $m_\chi$ and $m_G$, respectively. From Eq.~(\ref{eq:spin12_lag_full}) and using standard Feynman rules, we obtain the following amplitude for DM-nucleus scattering
\begin{equation}
\label{eq:spin12_matrixelement}
i\mathcal{M} = -\frac{i}{m_{G}^2}\ubar_\chi(p',s')\gamma^\mu(\lambda_3 + \lambda_4\gamma_5)u_\chi(p,s)\ubar_N(k',r')\gamma_\mu(h_3 + h_4\gamma_5)u_N(k,r)\,,
\end{equation}
where the four component spinors $u_\chi$ and $\ubar_\chi$ ($u_N$ and $\ubar_N$) are free spinor solutions to the Dirac equation for $\chi$ ($N$).~In the above expression, we assumed that the mediator particle is heavy, which implies $q^2\ll m_G^2$.~Here, we are interested in processes where a DM particle scatters off an initially polarised nucleus.~We therefore assume that the initial nuclear spin index $r$ is fixed, while $s, s', r' \in \{1,2\}$.~The spinor bilinears in Eq.~(\ref{eq:spin12_matrixelement}) admit the following non-relativistic expansion~\cite{DelNobile:2013sia}
\begin{align}
\label{eq:bilinar_list}
\ubar_\chi(p',s')\gamma^\mu u_\chi(p,s) &= \lvector{2m_\chi\delta^{s's}}{\vec{P}\delta^{s's} - 2i\vec{q}\times\sx},\nonumber \\
\ubar_\chi(p',s')\gamma^\mu\gamma_5 u_\chi(p,s) &= \lvector{2\vec{P}\cdot\sx}{4m_\chi\sx}, \nonumber\\
\ubar_N(k',r')\gamma_\mu u_N(k,r) &= \lvector{2m_N\delta^{r'r}}{-\vec{K}\delta^{r'r} - 2i\vec{q}\times\sn}, \nonumber\\
\ubar_N(k',r')\gamma_\mu\gamma_5 u_N(k,r) &= \lvector{2\vec{K}\cdot\sn}{-4m_N\sn} \,,
\end{align}
where $\vec{P} \equiv \vec{p} + \vec{p}\,'$,  $\vec{K} \equiv \vec{k} + \vec{k}\,'$, $\sn=\xi^{r' \dagger}(\vec{\sigma}_N/2)\xi^r$, $\sx=\xi^{s' \dagger}(\vec{\sigma}_\chi/2)\xi^s$, and $\vec{\sigma}_N$ and $\vec{\sigma}_\chi$ are tridimensional vectors whose components are Pauli matrices acting on the nucleus and DM particle spin space, respectively.~Here, $\xi^r$ and $\xi^s$ are two component spinors; the former arises from $u_N$ the latter from $u_\chi$.~From here onwards, we will denote by $\vec{s}$ the matrix element $\xi^{r \dagger }\,\vec{\sigma}_{N}\,\xi^r$ of the nuclear polarisation operator (therefore omitting the fixed polarisation index $r$ in the definition of $\vec{s}$).~Replacing Eq.~(\ref{eq:bilinar_list}) into Eq.~(\ref{eq:spin12_matrixelement}), we find the following expression for $i\mathcal{M}$
\begin{align}
\label{eq:spin12_result_original}
i\mathcal{M} &= -\frac{i}{m_{G}^2}\bigg\{\lambda_3h_3\bigg[4m_Nm_\chi\delta^{s's}\delta^{r'r}\bigg] \nonumber \\ 
&+ \lambda_3h_4\bigg[-8m_Nm_\chi\delta^{s's}\vt\cdot\sn  + 8im_N\sx\cdot(\sn\times\vec{q})\bigg] \nonumber \\ 
&+ \lambda_4h_3\bigg[8m_Nm_\chi\delta^{r'r}\vt\cdot\sx + 8im_\chi\sx\cdot(\sn\times\vec{q})\bigg] \nonumber \\ 
&+ \lambda_4h_4\bigg[-16m_Nm_\chi\sx\cdot\sn\bigg]\bigg\} \,,
\end{align}
where we only considered terms at most linear in $\vec{q}$ and $\vt$.~Now we calculate $|\overline{\niM}|^2$, the squared modulus of $i\mathcal{M}$ averaged over initial DM spins, and summed over final DM and nuclear spin states,
\begin{equation}
 |\overline{\niM}|^2 \equiv \frac{1}{2}\sum_{ss'}\sum_{r'}|\niM|^2 \,.
\end{equation} 
From Eq.~(\ref{eq:spin12_result_original}), we find
\begin{align}
\label{eq:result1}
|\overline{\niM}|^2 & =\frac{16m_\chi^2m_N^2}{m_G^4}\Bigg\{\lambda_3^2h_3^2 + 3\lambda_4^2h_4^2 \nonumber \\
&- \left[\lambda_3^2h_3h_4\left(1 - \frac{m_\chi}{m_N}\right) + \lambda_4^2h_3h_4\left(1 + \frac{m_\chi}{m_N}\right) + 2\lambda_3\lambda_4h_4^2\right]\vec{v}\cdot\vec{s}  \nonumber\\
&- \left[\lambda_3^2h_3h_4\left(1 + \frac{m_\chi}{m_N}\right) + \lambda_4^2h_3h_4\left(1 - \frac{m_\chi}{m_N}\right) - 2\lambda_3\lambda_4h_4^2\right]\vec{v}\,'\cdot\vec{s}\Bigg\} \,,
\end{align}
where we eliminated $\vec{q}$ in favour of $\vec{v}\,'$ by using Eq.~(\ref{eq:vt}), considered terms in the non-relativistic expansion of $|\overline{\niM}|^2$ which are at most linear in $\vec{v}$ and $\vec{v}\,'$, and applied the following spin summation rules
\begin{align}
\sum_{r'}\sndag\cdot\sn &= \frac{3}{4} \,, \nonumber\\
\sum_{r'}\sndag\times\sn &= \frac{i}{2}\vec{s} \,, \nonumber \\ 
\sum_{ss'}(\vec{a}\cdot\sxdag)(\vec{b}\cdot\sx) &= \frac{1}{2}\vec{a}\cdot \vec{b} \,,
\end{align}
where $\vec{a}$ and $\vec{b}$ are arbitrary tridimensional vectors.~Eq.~(\ref{eq:result1}) can be rewritten in a more compact form
\begin{align}
\label{eq:result1_again_again}
|\overline{\niM}|^2 = \frac{16m_\chi^2m_N^2}{m_G^4}&\bigg\{A - B\vec{v}\cdot\vec{s} - C\vec{v}\,'\cdot\vec{s}\bigg\} \,,
\end{align}
if we define
\begin{align}
A &\equiv \lambda_3^2h_3^2 + 3\lambda_4^2h_4^2\,,  \nonumber\\
B &\equiv \lambda_3^2h_3h_4\left(1 - \frac{m_\chi}{m_N}\right) + \lambda_4^2h_3h_4\left(1 + \frac{m_\chi}{m_N}\right) + 2\lambda_3\lambda_4h_4^2\,, \nonumber\\
C &\equiv \lambda_3^2h_3h_4\left(1 + \frac{m_\chi}{m_N}\right) + \lambda_4^2h_3h_4\left(1 - \frac{m_\chi}{m_N}\right) - 2\lambda_3\lambda_4h_4^2\,.
\end{align}
By replacing Eq.~(\ref{eq:result1}) into Eqs.~(\ref{eq:diff_cross_section1}) and (\ref{eq:triple_diff_event_rate_spherical3}), we are now able to numerically evaluate differential cross section and event rate for polarised DM-nucleus scattering in the case of spin 1/2 DM.

\subsection{Vector dark matter}
In the case of vector DM, we calculate $|\overline{\niM}|^2$ assuming the following Lagrangian for DM-nucleus interactions
\begin{align}
\label{eq:spin1_lag_full}
\mathscr{L} &=-\frac{1}{2}\mathcal{X}^\dagger_{\mu\nu}\mathcal{X}^{\mu\nu} + m_{X}^2X^\dagger_\mu X^\mu - \frac{\lambda_X}{2}\left(X^\dagger_\mu X^\mu\right)^2 \nonumber \\
&- \frac{1}{4}\mathcal{G}_{\mu\nu}\mathcal{G}^{\mu\nu} + \frac{1}{2}m_G^2G_\mu^2 - \frac{\lambda_G}{2}\Big(G_\mu G^\mu\Big)^2 \nonumber\\
&+ i\bar{N}\slashed DN - m_{N}\bar{N}N \nonumber\\
&-\frac{b_3}{2}G_\mu^2\left(X^\dagger_\nu X^\nu\right) - \frac{b_4}{2}\Big(G^\mu G^\nu\Big)\left(X^\dagger_\mu X_\nu\right) \nonumber\\
&- ib_5\left(X_\nu^\dagger\partial_\mu X^\nu - (\partial_\mu X^{\dagger\nu})X_\nu \right)G^\mu\nonumber\\
&- b_6 X^\dagger_\mu(\partial^\mu X_\nu) G^\nu - b_6^* (\partial^\mu X^\dagger_\nu)X_\mu G^\nu \nonumber\\
&- b_7\epsilon_{\mu\nu\rho\sigma}\left(X^{\dagger\mu}\partial^\nu X^\rho\right)G^\sigma - b_7^*\epsilon_{\mu\nu\rho\sigma}\left((\partial^\nu X^{\dagger\rho})X^{\mu}\right)G^\sigma \nonumber\\
&- h_3G_\mu \bar{N}\gamma^\mu N - h_4G_\mu\bar{N}\gamma^\mu\gamma_5N \,,
\end{align}
where $X_\mu$ is the DM vector field, $\mathcal{X}_{\mu\nu}$ and $m_X$ the associated field strength tensor and mass, respectively, and the other fields have the same meaning as in Eq.~(\ref{eq:spin12_lag_full}).~In Eq.~(\ref{eq:spin1_lag_full}), $b_6$ and $b_7$ are complex coupling constants, whereas without loss of generality, $b_3$, $b_4$ and $b_5$ can be taken as real.~It is convenient to rewrite $\mathscr{L}_{\rm int}$, the part involving $b_6$ and $b_7$ in the above Lagrangian, as follows
\begin{align}
\label{eq:spin1_lagrangian_expanded}
\mathscr{L}_{\rm int} & = - \Re(b_6)\partial_\nu\left( X^{\dagger\nu}X_\mu + X^\dagger_\mu X^\nu\right) G^\mu - i\Im(b_6)\partial_\nu\left( X^{\dagger\nu}X_\mu - X^\dagger_\mu X^\nu\right) G^\mu \nonumber\\
&- \Re(b_7)\epsilon_{\mu\nu\rho\sigma}\left(X^{\dagger\mu}\partial^\nu X^\rho + X^{\mu}\partial^\nu X^{\dagger\rho}\right)G^\sigma\nonumber\\
&- i\Im(b_7)\epsilon_{\mu\nu\rho\sigma}\left(X^{\dagger\mu}\partial^\nu X^\rho - X^{\mu}\partial^\nu X^{\dagger\rho}\right)G^\sigma \,.\end{align}
From Eqs.~(\ref{eq:spin1_lag_full}) and (\ref{eq:spin1_lagrangian_expanded}) and the Feynman rules in Appendix~\ref{sec:rules}, we find the DM-nucleus scattering amplitude 
\begin{align}
\label{eq:amplitude_spin1}
i\mathcal{M}&=-\frac{i b_5}{m_G^2} \left(p^{\prime\mu}+ p^\mu\right)\,\ubar_N(k',r')\gamma_\mu (h_3+h_4\gamma_5)u_N(k,r) \nonumber\\
&+\frac{\Re(b_6)}{m_G^2}\left(p_\nu' - p_\nu\right)\left(\epsilon^{s'\nu*}(p\,')\epsilon^{s\mu}(p)+\epsilon^{s'\mu*}(p\,')\epsilon^{s\nu}(p)\right) \,\ubar_N(k',r')\gamma_\mu (h_3+h_4\gamma_5)u_N(k,r) \nonumber\\
&+\frac{i\Im(b_6)}{m_G^2}\left(p_\nu' - p_\nu\right)\left(\epsilon^{s'\nu*}(p\,')\epsilon^{s\mu}(p)-\epsilon^{s'\mu*}(p\,')\epsilon^{s\nu}(p)\right)\,\ubar_N(k',r')\gamma_\mu(h_3+h_4\gamma_5) u_N(k,r) \nonumber\\
&-\frac{\Re(b_7)}{m_G^2}\varepsilon_{\mu\nu\rho\sigma}\left(p^{\prime\nu}+ p^\nu\right)\epsilon^{s'\mu*}(p\,')\epsilon^{s\rho}(p) \,\ubar_N(k',r')\gamma^{\sigma} (h_3+h_4\gamma_5)u_N(k,r) \nonumber\\
&+\frac{i\Im(b_7)}{m_G^2}\varepsilon_{\mu\nu\rho\sigma}\left(p^{\prime\nu}- p^\nu\right)\epsilon^{s'\mu*}(p\,')\epsilon^{s\rho}(p) \,\ubar_N(k',r')\gamma^\sigma (h_3+h_4\gamma_5)u_N(k,r) \,,
\end{align}
where $\epsilon^{s\mu}(p)$ and $\epsilon^{s'\nu*}(p\,')$ are polarisation vectors for the incoming and outgoing DM particle, respectively.~In a reference frame where the $z$-axis is in the direction of the DM particle momentum, they read 
\begin{align}
\label{eq:def_pol_states}
\epsilon^{s\mu}_X(p) = \begin{pmatrix}\dfrac{|\vec{p}\,|}{m_X}\delta^{3s}\\[5mm]
\dfrac{p_0}{m_\chi}\vec{e}_{s}\end{pmatrix}\,; \qquad\epsilon_X^{s'\mu*}(p') = \begin{pmatrix}\dfrac{|\vec{p}\,'|}{m_X}\delta^{3s'}\\[5mm]
\dfrac{p'_0}{m_\chi}\vec{e}\,'_{s'}\end{pmatrix}\,,
\end{align}
where $\vec{e}_3=\vec{p}/|\vec{p}|$ and $\vec{e}\,'_{3'}=\vec{p}\,'/|\vec{p}\,'|$.~In the non-relativistic limit, the scattering amplitude in Eq.~(\ref{eq:amplitude_spin1}) is given by
\begin{align}
\label{eq:amplitude_spin1_nonrel}
i\mathcal{M}&= -\frac{4ib_5h_3m_X m_N}{m_G^2}\delta^{r'r}\delta^{s's} + \frac{8ib_5h_4m_N m_X}{m_G^2}\vt\cdot\sn\delta^{s's}\nonumber\\
&-\frac{4\Re(b_6)h_3}{m_G^2}\bigg\{m_N\vec{q}\cdot\Sy^{s's}\cdot\vt\delta^{r'r} - i\left[\vec{q}\cdot\Sy^{s's}\cdot(\vec{q}\times\sn)\right]\bigg\} \nonumber\\
&+\frac{8\Re(b_6)h_4m_N}{m_G^2}\left(\vec{q}\cdot\Sy^{s's}\cdot\sn\right) \nonumber \\
&-\frac{i\Im(b_6)h_3}{m_G^2}\bigg\{\frac{2m_N}{m_X}\left(\vec{q}\cdot\Sy^{s's}\cdot\vec{q}\right)\delta^{r'r} +2im_N\sX\cdot(\vec{q}\times\vt)\delta^{r'r}\nonumber\\
&-2q^2(\sX\cdot\sn) + 2(\vec{q}\cdot\sX)(\vec{q}\cdot\sn)\bigg\} \nonumber\\
&-\frac{4\Im(b_6)h_4m_N}{m_G^2}(\vec{q}\times\sn)\cdot\sX \nonumber\\
 &+\frac{4i\Re(b_7)h_3m_X}{m_G^2}\bigg\{m_N\vt\cdot\sX\delta^{r'r} - i\sX\cdot(\vec{q}\times\sn)\bigg\} \nonumber\\
 &-\frac{8i\Re(b_7)h_4m_Nm_X}{m_G^2}\Big(\sX\cdot\sn\Big) \nonumber\\
 &+\frac{2\Im(b_7)h_3m_N}{m_G^2}\sX\cdot\vec{q}\,\delta^{r'r} \nonumber\\
 &+\frac{i\Im(b_7)h_4}{m_G^2}\bigg\{4im_N\left(\vec{q}\cdot\sX\right)\left(\vt\cdot\sn\right) + 4\frac{m_N}{m_X}\vec{q}\cdot\Sy^{s's}\cdot(\vec{q}\times\sn)\bigg\} \,,
\end{align}
where we only considered terms at most quadratic in $\vec{q}$ and linear in $\vt$, and, in analogy with Eq.~(\ref{eq:spin12_result_original}), introduced the notation
\begin{align}
\label{eq:spin1S}
\sX &= -i\esp\times\es\,,\nonumber\\
\Sy^{s's}_{ij} &= \frac{1}{2}\left(e_{si}e'_{s'j} + e_{sj}e'_{s'i}\right)\,.
\end{align}
In Eq.~(\ref{eq:spin1S}), $\Sy^{s's}_{ij}$ and $\sX$ are a symmetric and antisymmetric combination of polarisation vectors, respectively.~In particular, $\sX$ is a matrix element of the spin operator for spin 1 DM.~Finally, we calculate 
\begin{equation}
 |\overline{\niM}|^2 \equiv \frac{1}{3}\sum_{ss'}\sum_{r'}|\niM|^2 \,,
\end{equation} 
where we average over the three possible spin states of the incoming DM particle.~We find the following result
\begin{align}
\label{eq:result2}
|\overline{\niM}|^2 = \frac{16m_N^2m_X^2}{m_G^4}&\bigg\{I - J\vec{v}\cdot\vec{s} - K\vec{v}\,'\cdot\vec{s}\bigg\}\,,
\end{align}
where we only considered terms at most linear in $\vec{v}$ and $\vec{v}\,'$, and introduced the following coefficients
\begin{align}
I &\equiv b_5^2h_3^2 + 2\Re(b_7)^2h_4^2\,,\nonumber\\
J &\equiv \left(b_5^2h_3h_4 + \frac{2}{3}\Re(b_7)^2h_3h_4\right)\left(1 - \frac{m_X}{m_N}\right) - \frac{4}{3}\Re(b_7)h_4\left(\Im(b_6)h_4 - \frac{m_X}{m_N}\Re(b_7)h_3\right) \,,\nonumber\\
K &\equiv \left(b_5^2h_3h_4 + \frac{2}{3}\Re(b_7)^2h_3h_4\right)\left(1 + \frac{m_X}{m_N}\right)+ \frac{4}{3}\Re(b_7)h_4\left(\Im(b_6)h_4 - \frac{m_X}{m_N}\Re(b_7)h_3\right) \,, \nonumber\\
\end{align}
in analogy with the case of fermionic DM.~Deriving Eq.~(\ref{eq:result2}), we used the following spin summation rules, e.g.
\begin{align}
\label{xspin1_rule}
\sum_{ss'}(\vec{a}\cdot\sXdag)(\vec{b}\cdot\sX) &= 2(\vec{a}\cdot \vec{b})\,, \nonumber \\
\sum_{ss'}\vec{a}\cdot\Sy^{s's}\cdot \vec{b} \,\delta^{s's}&= (\vec{a}\cdot \vec{b}) \,,
\end{align}
where $\vec{a}$ and $\vec{b}$ are arbitrary tridimensional vectors.~Using Eq.~(\ref{eq:result2}), we can now numerically evaluate differential cross section and event rate for polarised DM-nucleus scattering in the case of spin 1 DM.

\section{Recoil energy spectra and angular distributions}
\label{sec:results}
The results of Sec.~\ref{sec:amplitudes} allow us to calculate nuclear recoil energy spectra and angular distributions resulting from the scattering of DM particles by polarised nuclei in direct detection experiments.~Nuclear recoil energy spectra are proportional to the integral of the triple differential rate in Eq.~(\ref{eq:triple_diff_event_rate_spherical3}) over all nuclear recoil directions within the solid angle $\alpha \in [0,\pi]$ and $\beta\in[0,2\pi]$.~Angular distributions are proportional to the energy integral of the triple differential rate in Eq.~(\ref{eq:triple_diff_event_rate_spherical3}).~In order to highlight the impact of polarisation dependent effects on physical observables, and compare results for fermionic and vector DM, it is convenient to study the function
\begin{align}
\label{eq:DeltaR}
\frac{{\rm d}\Delta R}{{\rm d}E_{\text{R}}{\rm d}\Omega} \equiv \frac{1}{2}\left(\frac{{\rm d}R(\vec{s}\,)}{{\rm d}E_{\text{R}}{\rm d}\Omega} - \frac{{\rm d}R(-\vec{s}\,)}{{\rm d}E_{\text{R}}{\rm d}\Omega}\right)\,,
\end{align}
instead of the triple differential rate in Eq.~(\ref{eq:triple_diff_event_rate_spherical3}).~Indeed, only $\vec{s}$ dependent terms survive in the difference of Eq.~(\ref{eq:DeltaR}), where we wrote ${\rm d}R$ as a function of $\vec{s}$ only, and omitted the dependence of ${\rm d}R$ on other variables.~The function in Eq.~(\ref{eq:DeltaR}) is the purely polarisation dependent differential scattering rate.~To numerically evaluate the latter for fermionic and vector DM, we must first set coupling constants and particle masses to benchmark values.~For the particle masses we choose $m_\chi=m_X=m_G=m_N=100$~GeV, since among the masses that can currently be probed at direct detection experiments.~Here we focus on the hypothetical case of a polarised detector employing heavy target nuclei.~For the coupling constants, we focus on a scenario where the unpolarised part of the scattering rate is the same for fermionic and vector DM.~This implies
\begin{align}
\label{eq:pcon}
\lambda_3^2h_3^2 + 3\lambda_4^2h_4^2 = b_5^2h_3^2 + 2\Re(b_7)^2h_4^2\,.
\end{align}
Furthermore, we assume maximal parity violation, in order to maximise the polarisation dependent contribution to the triple differential rate.~A possible choice of coupling constants compatible with these requirements is
\begin{align}
\label{eq:coeffs_choice}
&h_3 = \frac{1}{2}, &h_4 = -\frac{1}{2},& \quad \qquad \qquad \qquad \Im(b_6) = \frac{1}{2}, \nonumber \\
&\lambda_3 = \frac{1}{2}, &\lambda_4 = -\frac{1}{2},& \quad\qquad \qquad \qquad \Re(b_7) = -\frac{1}{2}, \nonumber \\
&b_5 = \frac{1}{\sqrt{2}}, &\Re(b_6)=0,& \quad\qquad \qquad \qquad \Im(b_7) = 0\,.
\end{align}
Notice that with this choice of coupling constants, the gamma matrices in the nuclear currents combine into left-handed projectors 
\begin{equation}
h_3 + h_4\gamma_5 =\frac{1-\gamma_5}{2}\,; \qquad \lambda_3 + \lambda_4\gamma_5 = \frac{1-\gamma_5}{2}\,.
\end{equation}
As an aside comment, we also mention that the $\Im(b_6)$ parameter only enters the purely polarisation dependent part of the scattering rate.~Consequently, one might in principle use measurements of polarisation dependent effects at direct detection experiments to constrain this parameter effectively.

\begin{figure}[t]
\begin{center}
\includegraphics[scale=0.36]{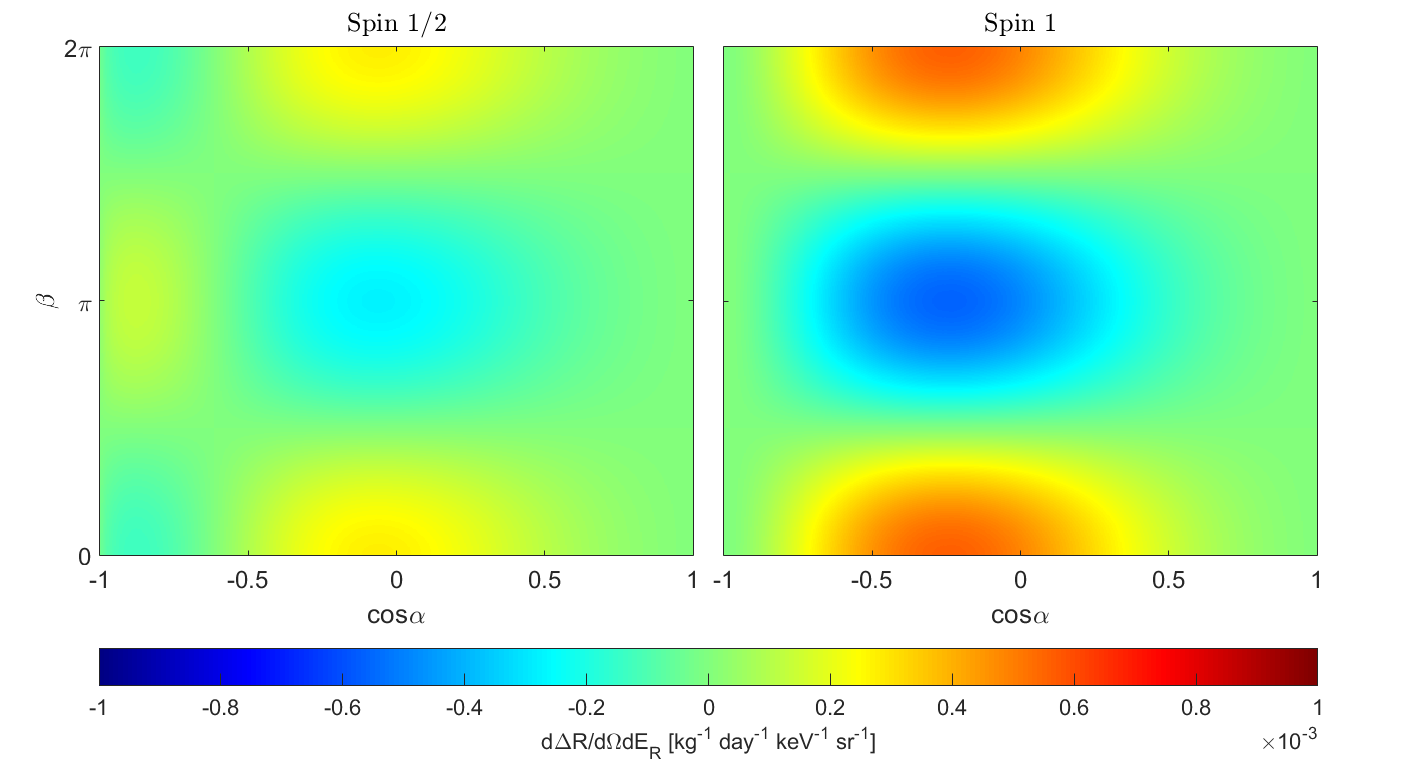}
\end{center}
\vspace{-0.45 cm}
\begin{center}
\hspace{0.12 cm}\includegraphics[scale=0.36]{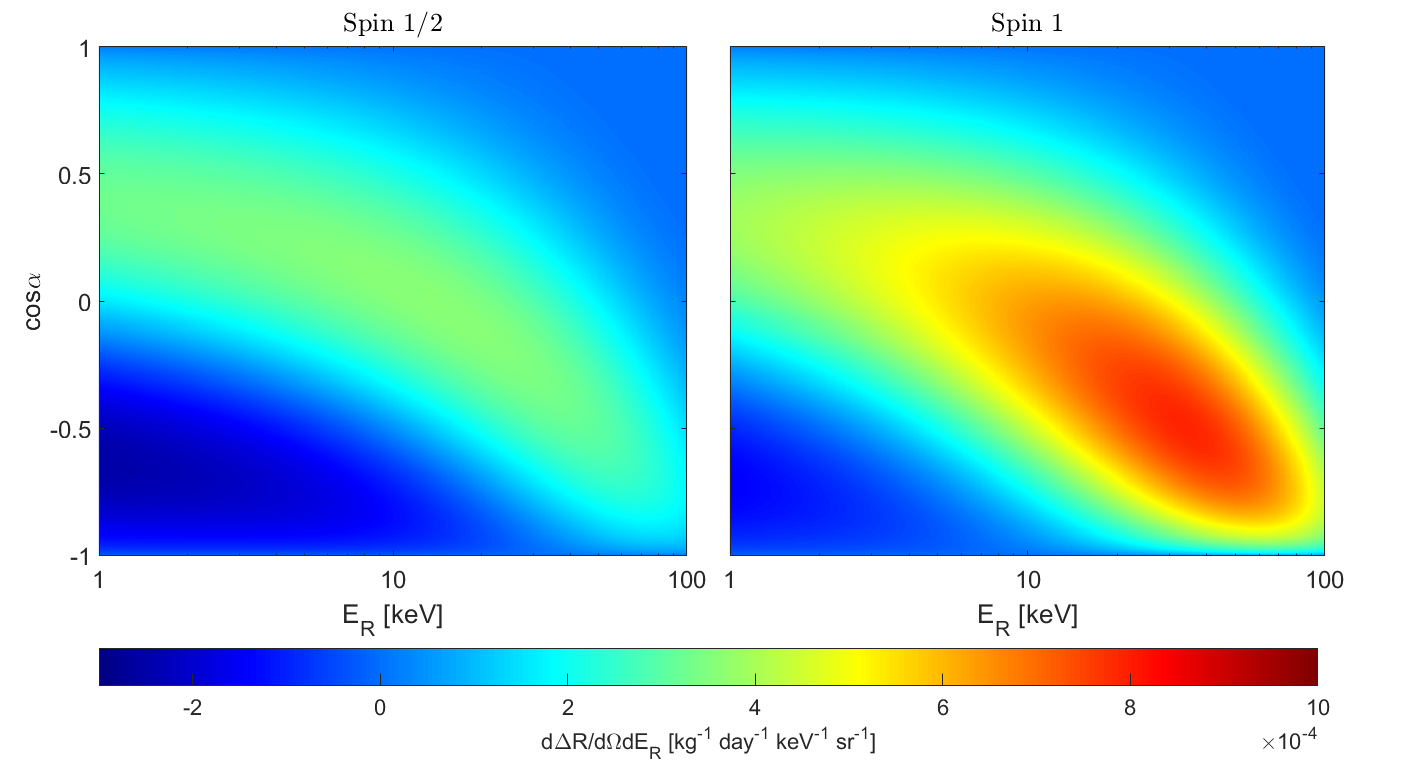}
\caption{\textbf{Top panels}.~Purely polarisation dependent part of the differential rate of scattering events plotted against the polar recoil angle $\alpha$ and the azimuthal angle $\beta$ for spin $1/2$ (left panel) and spin $1$ (right panel) DM.~The recoil energy $E_{\text{R}}$ is set to $20$~keV and the polarisation angle is fixed to $\vartheta = \pi/2$.~A nuclear recoil at $\cos\alpha = 0$, $\beta = 0$, would be in the direction of $\vec{s}$.~Model parameters have been set according to Eq.~(\ref{eq:coeffs_choice}).~\textbf{Bottom panels}.~Purely polarisation dependent part of the differential rate of scattering in the $E_R$, $\cos\alpha$ plane for spin $1/2$ (left panel) and spin $1$ (right panel) DM.~Here $\beta=0$ and $\vartheta=\pi/2$.}  
\label{fig:1}
\end{center}
\end{figure}

\begin{figure}[t]
\begin{center}
\includegraphics[scale=0.36]{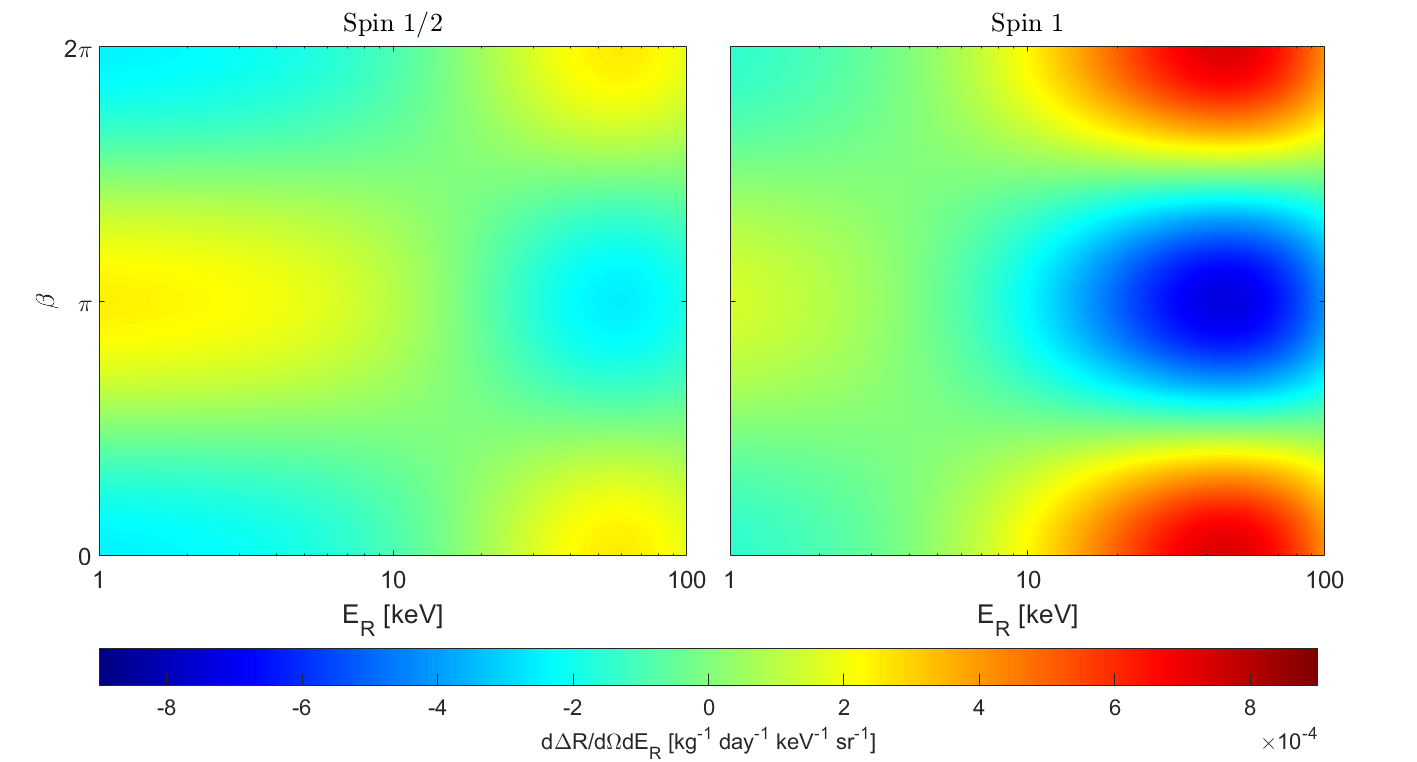}
\end{center}
\vspace{-0.1 cm}
\begin{center}
\hspace{0.12 cm}\includegraphics[scale=0.36]{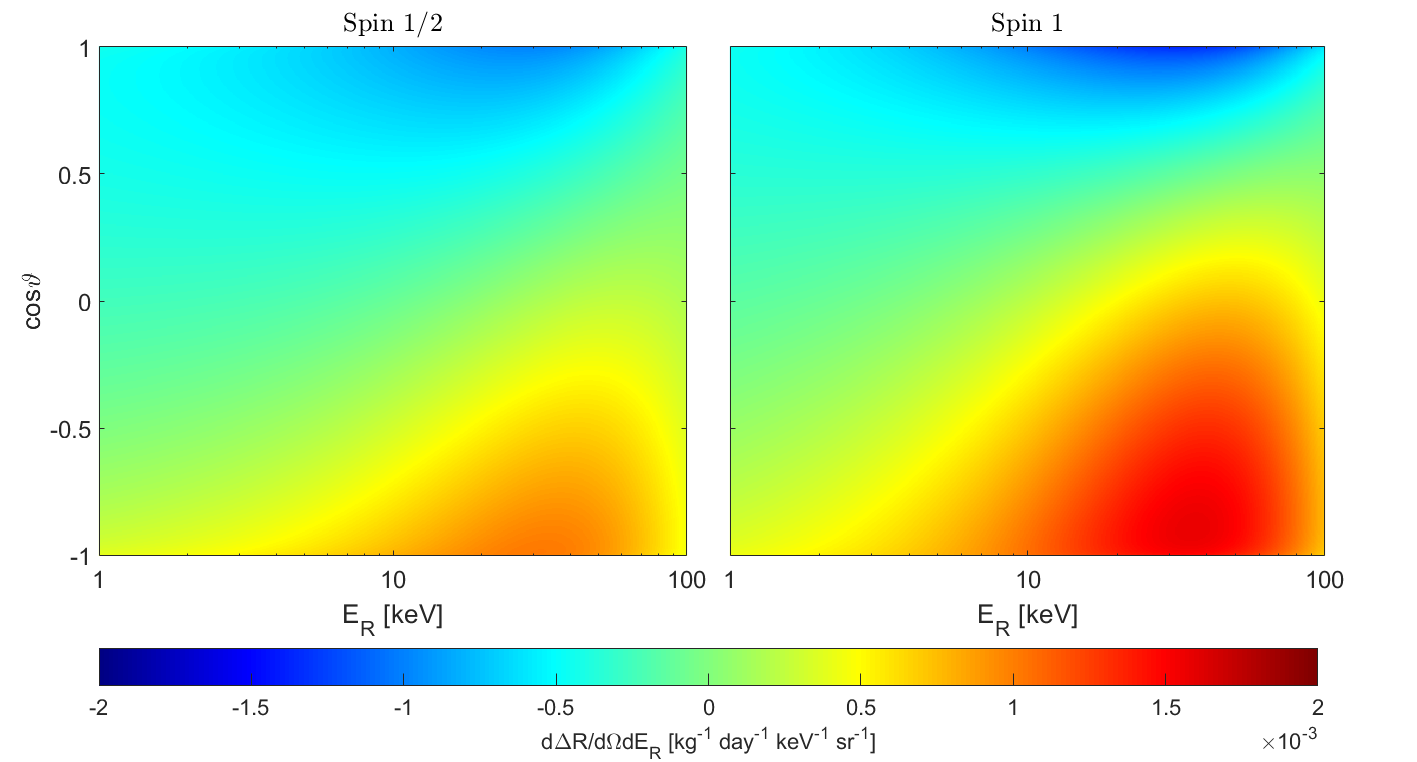}
\caption{\textbf{Top panels}.~Purely polarisation dependent part of the differential rate of scattering events in the $E_R$, $\beta$ plane for spin $1/2$ (left panel) and spin $1$ (right panel) DM.~The polar angle is set to $\alpha=3\pi/4$ and the polarisation angle is fixed to $\vartheta = \pi/2$.~Model parameters have been set according to Eq.~(\ref{eq:coeffs_choice}).~\textbf{Bottom panels}.~Purely polarisation dependent part of the differential rate of scattering events in the $E_R$, $\cos\vartheta$ plane for spin $1/2$ (left panel) and spin $1$ (right panel) DM.~Here, we assumed $\alpha=3\pi/4$ and $\beta=0$.}  
\label{fig:2}
\end{center}
\end{figure}

\begin{figure}[t]
\begin{center}
\includegraphics[scale=0.4]{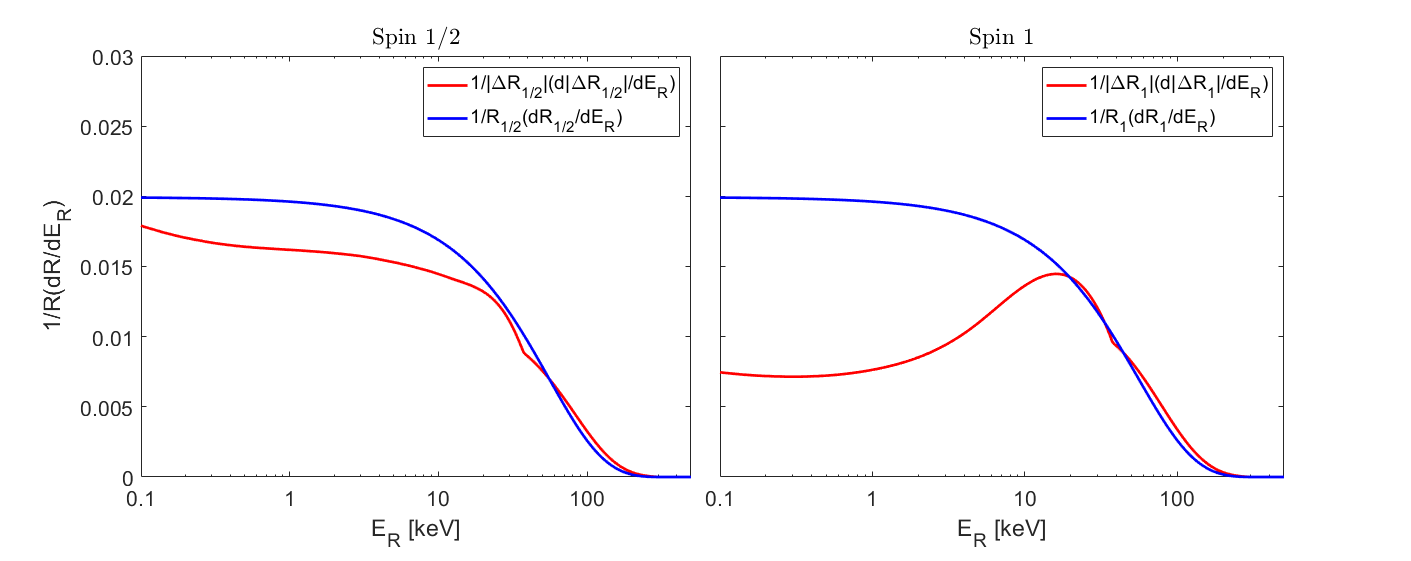}
\caption{Triple differential event rate, Eq.~(\ref{eq:triple_diff_event_rate_spherical3}), and its purely polarisation dependent part, Eq.~(\ref{eq:DeltaR}), integrated over all nuclear recoil angles, $\alpha$ and $\beta$, as a function of the nuclear recoil energy.~The left panel refers to fermionic DM, whereas the right panel corresponds to vector DM.~Each panel shows a blue line associated with the total rate, and a red line corresponding to its polarisation dependent part.}  
\label{fig:3}
\end{center}
\end{figure}
Let us now describe the results illustrated in Figs.~\ref{fig:1} and \ref{fig:2}.~In these figures, left panels refer to spin 1/2 DM, whereas right panels correspond to spin 1 DM.~Fig.~\ref{fig:1}, top panels, shows the purely polarisation dependent part of the scattering rate in the $\cos\alpha$, $\beta$ plane for a fixed recoil energy of 20 keV.~The polarisation angle is chosen to be $\vartheta = \pi/2$, which implies that a recoil at $\cos\alpha = 0$, $\beta = 0$, would be in the direction of the polarisation vector $\vec{s}$.~The azimuthal angle dependence in these plots is especially interesting, as it would not arise without parity violation, and, therefore, without a preferred direction in the problem, i.e.~$\vec{s}$, which breaks azimuthal symmetry.~In Fig.~\ref{fig:1} (top panels), we find a local maximum in the $\beta = 0$ direction at $\alpha=\alpha^*>\pi/2$ (evident in the case of spin 1 DM).~The exact value of $\alpha^*$ varies with recoil energy, as can be seen in Fig.~\ref{fig:1}, bottom panels, which shows Eq.~(\ref{eq:DeltaR}) at $\beta=0$ in the $E_R$, $\cos\alpha$ plane.~Not only the value of the azimuthal angle $\alpha$ at which Eq.~(\ref{eq:DeltaR}) is maximum depends on energy.~Fig.~\ref{fig:2}, top panels, shows the purely polarisation dependent part of the differential rate of scattering events at $\alpha=3\pi/4$ in the $E_R$, $\beta$ plane.~As one can see from this figure, the value of the azimuthal angle $\beta$ at which Eq.~(\ref{eq:DeltaR}) is maximum also depends on nuclear recoil energy.~It is maximum in a direction opposite to $\vec{s}$ for low energies, while for high energies it is in the same direction as the polarisation vector.~This leads to a certain critical energy where the $\beta$-dependence of the differential event rate in Eq.~(\ref{eq:DeltaR}) vanishes (around 10 keV for spin $1/2$ DM and 4 keV for spin $1$ DM). This critical energy depends on our choice of $\alpha$ ($\alpha = 3\pi/4$ in this case), and also on the choice of polarisation angle (here $\vartheta = \pi/2$).~The bottom panels in Fig.~\ref{fig:2} show the rate in Eq.~(\ref{eq:DeltaR}) at $\alpha = 3\pi/4$ and $\beta=0$ in the $E_R$, $\cos\vartheta$ plane.~For this choice of angles, and within the energy range considered here, the rate in Eq.~(\ref{eq:DeltaR}) is maximum when the spin of the target nuclei points in a direction opposite to the Earth's motion in the galactic rest frame.~It is minimum, when $\vec{v}_{\rm e}$ and $\vec{s}$ are aligned.

Fig.~\ref{fig:3} shows the triple differential event rate in Eq.~(\ref{eq:triple_diff_event_rate_spherical3}), and its purely polarisation dependent part, Eq.~(\ref{eq:DeltaR}), integrated over all nuclear recoil angles, $\alpha$ and $\beta$.~In the figure, rates are unit normalised, and $R_j$ ($\Delta R_j$), $j=1/2,1$, denotes Eq.~(\ref{eq:triple_diff_event_rate_spherical3}) (Eq.~(\ref{eq:DeltaR})) integrated over all energies and directions, with $j=1/2$ for spin 1/2 DM, and $j=1$ for spin 1 DM.~In Fig.~\ref{fig:3}, the left panel refers to fermionic DM, whereas the right panel corresponds to vector DM.~Each panel shows a blue line associated with the total rate, and a red line corresponding to its polarisation dependent part.~By construction, the two lines are independent of the recoil direction.~This figure illustrates an important result:~even without directional information (but assuming a large exposure), it could be possible to observe polarisation dependent effects in the DM-nucleus scattering.~Polarisation dependent effects are responsible for the differences observed in Fig.~\ref{fig:3} (left and right panels) between red and blue lines.~Such differences could be exploited to statistically discriminate between different DM particle spins.~For spin 1/2 DM, the purely polarisation dependent part of the scattering rate has a maximum in the small nuclear recoil energy limit, while it peaks at around 20 keV for spin $1$ DM.~Notice that, consistently with our choice of coupling constants, see Eq.~(\ref{eq:coeffs_choice}), the blue lines in the left and right panels of Fig.~\ref{fig:3} coincide.~On the other hand, the red lines in the two panels of Fig.~\ref{fig:3} are different, and their exact shape depends on our choice of parameters.~Consequently, if the polarisation dependent part of the DM-nucleus scattering rate were experimentally measurable, it would provide us with further information on the underlying Lagrangian for DM-nucleus interactions.

\section{Conclusion}
\label{sec:conclusion}
We studied the scattering of DM particles by spin-polarised target nuclei within a general set of simplified models for fermionic and vector DM, assuming that the scattering is mediated by a vector or pseudo-vector particle.~This study was motivated by the possibility of using polarised targets to gain novel insights into the nature of DM which would otherwise be difficult to extract from the result of operating direct detection experiments.~We presented our results in terms of nuclear recoil energy spectra and angular distributions, and provided explicit expressions for the underlying scattering amplitudes and cross-sections, separately focusing on spin 1/2 and spin 1 DM.~In the case of spin 1/2 DM, we refined previous results in the literature by correcting two sign errors in the corresponding cross section for polarised DM-nucleus scattering.~The new expression presented here is validated through a second calculation based on a different method, and by showing that the two results coincide.~In the case of spin 1 DM, we calculated the cross section for polarised DM-nucleus scattering for the first time.

We found that a signal at polarised direct detection experiments could in principle be used to discriminate fermionic from vector DM.~While this result is potentially very important, either a large exposure, or both directional sensitivity and the use of polarised targets is required in order to measure polarisation dependent effects in the rate of DM-nucleus scattering.~Roughly, since $\Delta R_j/R_j \sim 10^{-3}$, about $3\times10^6$ events would be necessary for a $3\sigma$ detection of a polarisation dependent signal at polarised direct detection experiments without directional sensitivity.~This number of events is comparable with the total number of nuclear recoils observed in DAMA~\cite{Chiang:2012ze}.~On the other hand, the required number of signal events is expected to decrease significantly when directional information is available.~A quantitative answer to these questions would require Monte Carlo simulations, and goes beyond the scope of this article.~However, the equations derived in this work set the ground for a critical assessment of the prospects for DM (spin) identification at next generation polarised DM direct detection experiments.

\acknowledgments It is a great pleasure to thank Anton B\"ackstr\"om, Anastasia Danopoulou, Fredrik Hellstr\"om and Martin B.~Krauss for useful and interesting discussions on dark matter direct detection.~We are also grateful to Stephon Alexander, Leah Jenks and Evan McDonough for finding an error in a first version of Eq.~(\ref{eq:triple_diff_event_rate_spherical3}).~This work was supported by the Knut and Alice Wallenberg Foundation (PI:~Jan Conrad) and is partly performed within the Swedish Consortium for Dark Matter Direct Detection (SweDCube).

\appendix

\section{Feynman rules}
\label{sec:rules}
Here we list some of the Feynman rules used in our calculations for vector DM.~We start from the interaction Lagrangian
\begin{align}
\mathscr{L}_{\rm int} = - ib_5\left(X_\nu^\dagger\partial_\mu X^\nu - (\partial_\mu X^{\dagger\nu})X_\nu \right)G^\mu
\end{align}
which corresponds to an interaction vertex with two DM particles and one mediator.~The associated Feynman rule is
\begin{align}
\feynmandiagram[large, horizontal =b to d,baseline = (b.base)]{
a[particle=\((\mu)\)]--[boson, momentum=\(q\), edge label'=\(G\)]
b,c[particle=\((\nu)\)]--[boson, momentum=\(p\), edge label'=\(X\)]
b,b--[boson, momentum=\(p'\), edge label'=\(X\)]
d[particle=\((\nu)\)]}; 
= -i b_5 \left(p_\mu' + p_\mu\right) \,.
\label{eq:b5_boson_vertex}
\end{align}
Similarly, the interaction Lagrangian 
\begin{align}
\mathscr{L}_{\rm int} &= - \Re(b_6)\partial_\nu\left( X^{\dagger\nu}X_\mu + X^\dagger_\mu X^\nu\right) G^\mu \nonumber\\ &- i\Im(b_6)\partial_\nu\left( X^{\dagger\nu}X_\mu - X^\dagger_\mu X^\nu\right) G^\mu 
\end{align}
is associated with the following interaction vertexes and Feynman rules
\begin{align}
&\feynmandiagram[large, horizontal =b to d,baseline = (b.base)]{
a[particle=\((\alpha)\)]--[boson, momentum=\(q\), edge label'=\(G\)]
b,c[particle=\((\gamma)\)]--[boson, momentum=\(p\), edge label'=\(X\)]
b,b--[boson, momentum=\(p'\), edge label'=\(X\)]
d[particle=\((\beta)\)]}; 
= \Re(b_6)\left(p_\nu' - p_\nu\right)\Big(\eta^{\nu\beta}\delta^\gamma_\mu+\delta^\beta_\mu\eta^{\nu\gamma}\Big)\eta^{\mu\alpha}
\nonumber\\
&\feynmandiagram[large, horizontal =b to d,baseline = (b.base)]{
a[particle=\((\alpha)\)]--[boson, momentum=\(q\), edge label'=\(G\)]
b,c[particle=\((\gamma)\)]--[boson, momentum=\(p\), edge label'=\(X\)]
b,b--[boson, momentum=\(p'\), edge label'=\(X\)]
d[particle=\((\beta)\)]}; 
= i\Im(b_6)\left(p_\nu' - p_\nu\right)\Big(\eta^{\nu\beta}\delta^\gamma_\mu-\delta^\beta_\mu\eta^{\nu\gamma}\Big)\eta^{\mu\alpha} \,.
\label{eq:b6_boson_vertex}
\end{align}
Finally, from the interaction Lagrangian
\begin{align}
\mathscr{L}_{\rm int} &= - \Re(b_7)\epsilon_{\mu\nu\rho\sigma}\left(X^{\dagger\mu}\partial^\nu X^\rho + X^{\mu}\partial^\nu X^{\dagger\rho}\right)G^\sigma \nonumber\\ &- i\Im(b_7)\epsilon_{\mu\nu\rho\sigma}\left(X^{\dagger\mu}\partial^\nu X^\rho - X^{\mu}\partial^\nu X^{\dagger\rho}\right)G^\sigma
\end{align}
we obtain the following interaction vertexes and Feynman rules
\begin{align}
&\feynmandiagram[large, horizontal =b to d,baseline = (b.base)]{
a[particle=\((\sigma)\)]--[boson, momentum=\(q\), edge label'=\(G\)]
b,c[particle=\((\rho)\)]--[boson, momentum=\(p\), edge label'=\(X\)]
b,b--[boson, momentum=\(p'\), edge label'=\(X\)]
d[particle=\((\mu)\)]}; 
= -\Re(b_7)\varepsilon_{\mu\nu\rho\sigma}\left(p^\nu + p^{\prime\nu}\right)
\nonumber\\
&\feynmandiagram[large, horizontal =b to d,baseline = (b.base)]{
a[particle=\((\sigma)\)]--[boson, momentum=\(q\), edge label'=\(G\)]
b,c[particle=\((\rho)\)]--[boson, momentum=\(p\), edge label'=\(X\)]
b,b--[boson, momentum=\(p'\), edge label'=\(X\)]
d[particle=\((\mu)\)]}; 
= -i\Im(b_7)\varepsilon_{\mu\nu\rho\sigma}\left(p^\nu - p^{\prime\nu}\right) \,.
\label{eq:b7_boson_vertex}
\end{align}

\section{Validation of Eq.~(\ref{eq:result1})}
\label{sec:vector_v2}
In order to validate Eq.~(\ref{eq:result1}), below we provide an independent derivation of the same expression.~As for the calculation in Sec.~\ref{sec:amplitudes}, the starting point for this second derivation is Eq.~(\ref{eq:spin12_matrixelement}).~Instead of taking the non-relativistic limit of Eq.~(\ref{eq:spin12_matrixelement}), and then computing the squared modulus of the obtained result, here we calculate the squared modulus of the relativistic expression in Eq.~(\ref{eq:spin12_matrixelement}), and only at the end of the calculation take the non-relativistic limit.~For the squared modulus of the relativistic expression for the amplitude $i\mathcal{M}$ in Eq.~(\ref{eq:spin12_matrixelement}), we find
\begin{align}
\label{eq:spin12_spinor_big}
|\overline{\niM}|^2  & = \frac{4}{m_{G}^4}\bigg[\lambda_3^2h_3^2\big(2p'\cdot k' p\cdot k + 2p'\cdot k p\cdot k' - 2m_\chi^2k'\cdot k - 2m_N^2p'\cdot p + 4m_N^2m_\chi^2\big) \nonumber \\
&+\lambda_3^2h_4^2\big(2p'\cdot k' p\cdot k + 2p'\cdot k p\cdot k' - 2m_\chi^2k'\cdot k + 2m_N^2p'\cdot p - 4m_N^2m_\chi^2\big) \nonumber\\
&+\lambda_4^2h_3^2\big(2p'\cdot k' p\cdot k + 2p'\cdot k p\cdot k' + 2m_\chi^2k'\cdot k - 2m_N^2p'\cdot p - 4m_N^2m_\chi^2\big) \nonumber\\
&+\lambda_4^2h_4^2\big(2p'\cdot k' p\cdot k + 2p'\cdot k p\cdot k' + 2m_\chi^2k'\cdot k + 2m_N^2p'\cdot p + 4m_N^2m_\chi^2\big) \nonumber\\
&-2\lambda_3^2h_3h_4\big(2m_N p'\cdot k' p \cdot S + 2m_N k'\cdot p p'\cdot S - 2m_\chi^2m_N k'\cdot S\big) \nonumber\\
&-2\lambda_4^2h_3h_4\big(2m_N p'\cdot k' p \cdot S + 2m_N k'\cdot p p'\cdot S + 2m_\chi^2m_N k'\cdot S\big) \nonumber\\
&-4\lambda_3\lambda_4h_3^2\big(m_Np'\cdot k' p \cdot S - m_N k'\cdot p p'\cdot S + m_N p\cdot k p'\cdot S - m_N p'\cdot k p \cdot S\big) \nonumber\\
&-4\lambda_3\lambda_4h_4^2\big(m_Np'\cdot k' p \cdot S - m_N k'\cdot p p'\cdot S - m_N p\cdot k p'\cdot S + m_N p'\cdot k p \cdot S\big) \nonumber\\
&-8\lambda_3\lambda_4h_3h_4\big(k'\cdot p' k\cdot p - k'\cdot p k \cdot p'\big)\bigg] \,,
\end{align}
where a dot denotes the scalar product between four-vectors, and the four-vector $S = (0,\vec{s})$ fulfils the identity~\cite{QFT}
\begin{equation}
\label{eq:srednicki_trick}
u_{N,a}(k,r)\ubar_{N,b}(k,r) = \frac{1}{2}(1 - \gamma_5\slashed S)_{ac}(\slashed k + m_N)_{cb}\,,
\end{equation}
where $a$, $b$ and $c$ are four-component spinor indeces.~In order to compare the above expression with Eq.~(\ref{eq:result1}), we must take the non-relativistic limit of Eq.~(\ref{eq:spin12_spinor_big}).~To this end, we replace the four-momenta in Eq.~(\ref{eq:spin12_spinor_big}) with $p = m_\chi(1, \vec{v}\,)$, $k = (m_N, 0)$, $p' = m_\chi(1, \vec{v}\,')$ and $k' = (m_N, -\vec{q}\,) $, where we used the notation of Sec.~\ref{sec:theory}, assumed the target nucleus to be initially at rest, and defined the momentum transfer as follows $\vec{q} = m_\chi(\vec{v}\,' - \vec{v}\,)$.~With this replacement we find
\begin{align} 
\label{eq:spin12_spinor_nonrel}
|\overline{\niM}|^2  &= \frac{8m_\chi^2m_N^2}{m_{G}^4}\Bigg\{2\lambda_3^2h_3^2 + 6\lambda_4^2h_4^2 - 2\lambda_3^2h_3h_4\Bigg[\left(1 - \frac{m_\chi}{m_N}\right)\vec{v}\cdot\vec{s} + \left(1 + \frac{m_\chi}{m_N}\right)\vec{v}\,'\cdot \vec{s}\,\Bigg] \nonumber\\
&- 2\lambda_4^2h_3h_4\Big[\left(1 + \frac{m_\chi}{m_N}\right)\vec{v}\cdot\vec{s} + \left(1 - \frac{m_\chi}{m_N}\right)\vec{v}\,'\cdot \vec{s}\,\Big]\nonumber\\
&- 4\lambda_3\lambda_4h_4^2\left(\vec{v}\cdot\vec{s}- \vec{v}\,'\cdot\vec{s}\,\right)\Bigg\} \,.
\end{align}
After collecting the terms proportional to $\vec{v}\,'\cdot\vec{s}$ and $\vec{v}\cdot\vec{s}$, the above expression coincides with Eq.~(\ref{eq:result1}).


\begin{thebibliography}{10}

\bibitem{Bertone:2004pz}
G.~Bertone, D.~Hooper and J.~Silk, \emph{{Particle dark matter: Evidence,
  candidates and constraints}},
  \href{https://doi.org/10.1016/j.physrep.2004.08.031}{\emph{Phys. Rept.}
  {\bfseries 405} (2005) 279}
  [\href{https://arxiv.org/abs/hep-ph/0404175}{{\ttfamily hep-ph/0404175}}].

\bibitem{Jungman:1995df}
G.~Jungman, M.~Kamionkowski and K.~Griest, \emph{{Supersymmetric dark matter}},
  \href{https://doi.org/10.1016/0370-1573(95)00058-5}{\emph{Phys. Rept.}
  {\bfseries 267} (1996) 195}
  [\href{https://arxiv.org/abs/hep-ph/9506380}{{\ttfamily hep-ph/9506380}}].

\bibitem{Bertone:2016nfn}
G.~Bertone and D.~Hooper, \emph{{A History of Dark Matter}}, {\emph{Submitted
  to: Rev. Mod. Phys.} (2016) }
  [\href{https://arxiv.org/abs/1605.04909}{{\ttfamily 1605.04909}}].

\bibitem{Arcadi:2017kky}
G.~Arcadi, M.~Dutra, P.~Ghosh, M.~Lindner, Y.~Mambrini, M.~Pierre et~al.,
  \emph{{The waning of the WIMP? A review of models, searches, and
  constraints}},
  \href{https://doi.org/10.1140/epjc/s10052-018-5662-y}{\emph{Eur. Phys. J.}
  {\bfseries C78} (2018) 203}
  [\href{https://arxiv.org/abs/1703.07364}{{\ttfamily 1703.07364}}].

\bibitem{Roszkowski:2017nbc}
L.~Roszkowski, E.~M. Sessolo and S.~Trojanowski, \emph{{WIMP dark matter
  candidates and searches - current status and future prospects}},
  \href{https://doi.org/10.1088/1361-6633/aab913}{\emph{Rept. Prog. Phys.}
  {\bfseries 81} (2018) 066201}
  [\href{https://arxiv.org/abs/1707.06277}{{\ttfamily 1707.06277}}].

\bibitem{Drukier:1983gj}
A.~Drukier and L.~Stodolsky, \emph{{Principles and Applications of a Neutral
  Current Detector for Neutrino Physics and Astronomy}},
  \href{https://doi.org/10.1103/PhysRevD.30.2295}{\emph{Phys. Rev.} {\bfseries
  D30} (1984) 2295}.

\bibitem{Goodman:1984dc}
M.~W. Goodman and E.~Witten, \emph{{Detectability of Certain Dark Matter
  Candidates}}, \href{https://doi.org/10.1103/PhysRevD.31.3059}{\emph{Phys.
  Rev.} {\bfseries D31} (1985) 3059}.

\bibitem{Baudis:2012ig}
L.~Baudis, \emph{{Direct dark matter detection: the next decade}},
  \href{https://doi.org/10.1016/j.dark.2012.10.006}{\emph{Phys. Dark Univ.}
  {\bfseries 1} (2012) 94} [\href{https://arxiv.org/abs/1211.7222}{{\ttfamily
  1211.7222}}].

\bibitem{Catena:2014epa}
R.~Catena, \emph{{Prospects for direct detection of dark matter in an effective
  theory approach}},
  \href{https://doi.org/10.1088/1475-7516/2014/07/055}{\emph{JCAP} {\bfseries
  1407} (2014) 055} [\href{https://arxiv.org/abs/1406.0524}{{\ttfamily
  1406.0524}}].

\bibitem{Catena:2014hla}
R.~Catena, \emph{{Analysis of the theoretical bias in dark matter direct
  detection}}, \href{https://doi.org/10.1088/1475-7516/2014/09/049}{\emph{JCAP}
  {\bfseries 1409} (2014) 049}
  [\href{https://arxiv.org/abs/1407.0127}{{\ttfamily 1407.0127}}].

\bibitem{Catena:2015vpa}
R.~Catena, \emph{{Dark matter directional detection in non-relativistic
  effective theories}},
  \href{https://doi.org/10.1088/1475-7516/2015/07/026}{\emph{JCAP} {\bfseries
  1507} (2015) 026} [\href{https://arxiv.org/abs/1505.06441}{{\ttfamily
  1505.06441}}].

\bibitem{Lewin:1995rx}
J.~Lewin and P.~Smith, \emph{{Review of mathematics, numerical factors, and
  corrections for dark matter experiments based on elastic nuclear recoil}},
  \href{https://doi.org/10.1016/S0927-6505(96)00047-3}{\emph{Astropart. Phys.}
  {\bfseries 6} (1996) 87}.

\bibitem{Undagoitia:2015gya}
T.~Marrodan~Undagoitia and L.~Rauch, \emph{{Dark matter direct-detection
  experiments}}, \href{https://doi.org/10.1088/0954-3899/43/1/013001}{\emph{J.
  Phys.} {\bfseries G43} (2016) 013001}
  [\href{https://arxiv.org/abs/1509.08767}{{\ttfamily 1509.08767}}].

\bibitem{Chiang:2012ze}
C.-T. Chiang, M.~Kamionkowski and G.~Z. Krnjaic, \emph{{Dark Matter Detection
  with Polarized Detectors}},
  \href{https://doi.org/10.1016/j.dark.2012.10.003}{\emph{Phys. Dark Univ.}
  {\bfseries 1} (2012) 109} [\href{https://arxiv.org/abs/1202.1807}{{\ttfamily
  1202.1807}}].

\bibitem{Franarin:2016ppr}
T.~Franarin and M.~Fairbairn, \emph{{Reducing the solar neutrino background in
  dark matter searches using polarized helium-3}},
  \href{https://doi.org/10.1103/PhysRevD.94.053004}{\emph{Phys. Rev.}
  {\bfseries D94} (2016) 053004}
  [\href{https://arxiv.org/abs/1605.08727}{{\ttfamily 1605.08727}}].

\bibitem{Amarian:2002ar}
M.~Amarian et~al., \emph{{The Q**2 evolution of the generalized
  Gerasimov-Drell-Hearn integral for the neutron using a He-3 target}},
  \href{https://doi.org/10.1103/PhysRevLett.89.242301}{\emph{Phys. Rev. Lett.}
  {\bfseries 89} (2002) 242301}
  [\href{https://arxiv.org/abs/nucl-ex/0205020}{{\ttfamily nucl-ex/0205020}}].

\bibitem{Amarian:2003jy}
{\scshape Jefferson Lab E94-010} collaboration, M.~Amarian et~al., \emph{{Q**2
  evolution of the neutron spin structure moments using a He-3 target}},
  \href{https://doi.org/10.1103/PhysRevLett.92.022301}{\emph{Phys. Rev. Lett.}
  {\bfseries 92} (2004) 022301}
  [\href{https://arxiv.org/abs/hep-ex/0310003}{{\ttfamily hep-ex/0310003}}].

\bibitem{Bouchiat:1960dsd}
M.~A. Bouchiat, T.~R. Carver and C.~M. Varnum, \emph{{Nuclear Polarization in
  He3 Gas Induced by Optical Pumping and Dipolar Exchange}},
  \href{https://doi.org/10.1103/PhysRevLett.5.373}{\emph{Phys. Rev. Lett.}
  {\bfseries 5} (1960) 373}.

\bibitem{Walker:1997zzc}
T.~G. Walker and W.~Happer, \emph{{Spin-exchange optical pumping of noble-gas
  nuclei}}, \href{https://doi.org/10.1103/RevModPhys.69.629}{\emph{Rev. Mod.
  Phys.} {\bfseries 69} (1997) 629}.

\bibitem{Brod:2017bsw}
J.~Brod, A.~Gootjes-Dreesbach, M.~Tammaro and J.~Zupan, \emph{{Effective Field
  Theory for Dark Matter Direct Detection up to Dimension Seven}},
  \href{https://arxiv.org/abs/1710.10218}{{\ttfamily 1710.10218}}.

\bibitem{Bishara:2016hek}
F.~Bishara, J.~Brod, B.~Grinstein and J.~Zupan, \emph{{Chiral Effective Theory
  of Dark Matter Direct Detection}},
  \href{https://doi.org/10.1088/1475-7516/2017/02/009}{\emph{JCAP} {\bfseries
  1702} (2017) 009} [\href{https://arxiv.org/abs/1611.00368}{{\ttfamily
  1611.00368}}].

\bibitem{Catena:2017xqq}
R.~Catena, J.~Conrad and M.~B. Krauss, \emph{{Compatibility of a dark matter
  discovery at XENONnT/LZ with the WIMP thermal production mechanism}},
  \href{https://doi.org/10.1103/PhysRevD.97.103002}{\emph{Phys. Rev.}
  {\bfseries D97} (2018) 103002}
  [\href{https://arxiv.org/abs/1712.07969}{{\ttfamily 1712.07969}}].

\bibitem{Baum:2017kfa}
S.~Baum, R.~Catena, J.~Conrad, K.~Freese and M.~B. Krauss, \emph{{Determining
  Dark Matter properties with a XENONnT/LZ signal and LHC-Run3 mono-jet
  searches}},  \href{https://arxiv.org/abs/1709.06051}{{\ttfamily 1709.06051}}.

\bibitem{Catena:2017wzu}
R.~Catena, J.~Conrad, C.~D{\"o}ring, A.~D. Ferella and M.~B. Krauss,
  \emph{{Dark matter spin determination with directional direct detection
  experiments}},  \href{https://arxiv.org/abs/1706.09471}{{\ttfamily
  1706.09471}}.

\bibitem{Dent:2015zpa}
J.~B. Dent, L.~M. Krauss, J.~L. Newstead and S.~Sabharwal, \emph{{General
  analysis of direct dark matter detection: From microphysics to observational
  signatures}}, \href{https://doi.org/10.1103/PhysRevD.92.063515}{\emph{Phys.
  Rev.} {\bfseries D92} (2015) 063515}
  [\href{https://arxiv.org/abs/1505.03117}{{\ttfamily 1505.03117}}].

\bibitem{Sivertsson:2017rkp}
S.~Sivertsson, H.~Silverwood, J.~I. Read, G.~Bertone and P.~Steger, \emph{{The
  Local Dark Matter Density from SDSS-SEGUE G-dwarfs}},
  \href{https://doi.org/10.1093/mnras/sty977}{\emph{Mon. Not. Roy. Astron.
  Soc.} {\bfseries 478} (2018) 1677}
  [\href{https://arxiv.org/abs/1708.07836}{{\ttfamily 1708.07836}}].

\bibitem{Bozorgnia:2013pua}
N.~Bozorgnia, R.~Catena and T.~Schwetz, \emph{{Anisotropic dark matter
  distribution functions and impact on WIMP direct detection}},
  \href{https://doi.org/10.1088/1475-7516/2013/12/050}{\emph{JCAP} {\bfseries
  1312} (2013) 050} [\href{https://arxiv.org/abs/1310.0468}{{\ttfamily
  1310.0468}}].

\bibitem{Catena:2011kv}
R.~Catena and P.~Ullio, \emph{{The local dark matter phase-space density and
  impact on WIMP direct detection}},
  \href{https://doi.org/10.1088/1475-7516/2012/05/005}{\emph{JCAP} {\bfseries
  1205} (2012) 005} [\href{https://arxiv.org/abs/1111.3556}{{\ttfamily
  1111.3556}}].

\bibitem{Catena:2009mf}
R.~Catena and P.~Ullio, \emph{{A novel determination of the local dark matter
  density}}, \href{https://doi.org/10.1088/1475-7516/2010/08/004}{\emph{JCAP}
  {\bfseries 1008} (2010) 004}
  [\href{https://arxiv.org/abs/0907.0018}{{\ttfamily 0907.0018}}].

\bibitem{Gondolo:2002np}
P.~Gondolo, \emph{{Recoil momentum spectrum in directional dark matter
  detectors}}, \href{https://doi.org/10.1103/PhysRevD.66.103513}{\emph{Phys.
  Rev.} {\bfseries D66} (2002) 103513}
  [\href{https://arxiv.org/abs/hep-ph/0209110}{{\ttfamily hep-ph/0209110}}].

\bibitem{DelNobile:2013sia}
M.~Cirelli, E.~Del~Nobile and P.~Panci, \emph{{Tools for model-independent
  bounds in direct dark matter searches}},
  \href{https://doi.org/10.1088/1475-7516/2013/10/019}{\emph{JCAP} {\bfseries
  1310} (2013) 019} [\href{https://arxiv.org/abs/1307.5955}{{\ttfamily
  1307.5955}}].

\bibitem{Catena:2014uqa}
R.~Catena and P.~Gondolo, \emph{{Global fits of the dark matter-nucleon
  effective interactions}},
  \href{https://doi.org/10.1088/1475-7516/2014/09/045}{\emph{JCAP} {\bfseries
  1409} (2014) 045} [\href{https://arxiv.org/abs/1405.2637}{{\ttfamily
  1405.2637}}].

\bibitem{Catena:2015uua}
R.~Catena and P.~Gondolo, \emph{{Global limits and interference patterns in
  dark matter direct detection}},
  \href{https://doi.org/10.1088/1475-7516/2015/08/022}{\emph{JCAP} {\bfseries
  1508} (2015) 022} [\href{https://arxiv.org/abs/1504.06554}{{\ttfamily
  1504.06554}}].

\bibitem{Lee:2013xxa}
S.~K. Lee, M.~Lisanti and B.~R. Safdi, \emph{{Dark-Matter Harmonics Beyond
  Annual Modulation}},
  \href{https://doi.org/10.1088/1475-7516/2013/11/033}{\emph{JCAP} {\bfseries
  1311} (2013) 033} [\href{https://arxiv.org/abs/1307.5323}{{\ttfamily
  1307.5323}}].

\bibitem{Freese:2012xd}
K.~Freese, M.~Lisanti and C.~Savage, \emph{{Colloquium: Annual modulation of
  dark matter}}, \href{https://doi.org/10.1103/RevModPhys.85.1561}{\emph{Rev.
  Mod. Phys.} {\bfseries 85} (2013) 1561}
  [\href{https://arxiv.org/abs/1209.3339}{{\ttfamily 1209.3339}}].

\bibitem{Fitzpatrick:2012ix}
A.~L. Fitzpatrick, W.~Haxton, E.~Katz, N.~Lubbers and Y.~Xu, \emph{{The
  Effective Field Theory of Dark Matter Direct Detection}},
  \href{https://doi.org/10.1088/1475-7516/2013/02/004}{\emph{JCAP} {\bfseries
  1302} (2013) 004} [\href{https://arxiv.org/abs/1203.3542}{{\ttfamily
  1203.3542}}].

\bibitem{Anand:2013yka}
N.~Anand, A.~L. Fitzpatrick and W.~Haxton, \emph{{Model-independent WIMP
  Scattering Responses and Event Rates: A Mathematica Package for Experimental
  Analysis}}, \href{https://doi.org/10.1103/PhysRevC.89.065501}{\emph{Phys.
  Rev.} {\bfseries C89} (2014) 065501}
  [\href{https://arxiv.org/abs/1308.6288}{{\ttfamily 1308.6288}}].

\bibitem{QFT}
M.~Srednicki, \emph{{Quantum Field Theory - Cambridge University Press}},
  2018.

\end{thebibliography}

\providecommand{\href}[2]{#2}\begingroup\raggedright\endgroup

\end{document}